# Retina electronic paper with video-rate-tunable 45,000 pixels per inch

**Authors:** Ade Satria Saloka Santosa[1], Yu-Wei Chang[2], Andreas B. Dahlin[3], Lars Österlund[1], Giovanni Volpe[2], Kunli Xiong[1]*

**Affiliations:**

[1]Department of Materials Science and Engineering, The Angstrom Laboratory, Uppsala University, SE-752 37 Uppsala, Sweden.

[2]Department of Physics, University of Gothenburg, SE-412 96 Gothenburg, Sweden.

[3]Department of Chemistry and Chemical Engineering, Chalmers University of Technology, SE-412 96 Gothenburg, Sweden.

*Corresponding author. Email: kunli.xiong@angstrom.uu.se

**Abstract:** As demand for immersive experiences grows, displays are moving closer to the eye with smaller sizes and higher resolutions. However, shrinking pixel emitters reduce intensity, making them harder to perceive. Electronic Papers utilize ambient light for visibility, maintaining optical contrast regardless of pixel size, but cannot achieve high resolution. We show electrically tunable meta-pixels down to ~560 nm in size (>45,000 PPI) consisting of $WO_3$ nanodiscs, allowing one-to-one pixel-photodetector mapping on the retina when the display size matches the pupil diameter, which we call **Retina Electronic Paper**. Our technology also supports video display (25 Hz), high reflectance (~80%), and optical contrast (~50%), which will help create the ultimate virtual reality display.

**Main Text:** From cinema screens and televisions to smartphones, and virtual reality (VR) headsets, displays have progressively moved closer to the human eye, featuring smaller sizes and higher resolutions. As display technology advances, a fundamental question arises: What are the ultimate limits of display size and resolution? As shown in Fig. 1A, to achieve the most immersive and optimal visual experience, the display should closely match the dimensions of the human pupil, with each pixel corresponding one-to-one with a photoreceptor cell in the retina. The human retina contains approximately 120 million photoreceptor cells. Assuming a pupil diameter of 8 mm, the ideal pixel size would be ~650 nm, resulting in a resolution of ~40,000 pixels per inch (PPI). This theoretical pixel size approaches the resolution limit of the human eye, representing the ultimate boundary for display technologies, which we name the "Retina" display.

Mainstream emissive displays are approaching their physical limits as pixel sizes shrink. A smaller pixel size reduces emitter dimensions, leading to a significant drop in luminosity, which makes them increasingly difficult to perceive with the naked eye (1, 2). Currently, commercially available smartphone display pixels are typically ~60 × 60 μm² (~450 PPI), ~10,000 times larger than the theoretical size required for the ultimate Retina display. Already at this scale, the emitted light becomes difficult for the naked eye to perceive, particularly in bright outdoor environments. Moreover, the smallest published micro-LEDs currently available only achieve a pixel size of 4 × 4 μm² (excluding the distance between pixels) (3), which is still two orders of magnitude larger than the area of the photoreceptors in the retina. These limitations expose the significant challenges of using conventional emissive display technology to realize the ultimate VR display.

Reflective displays, which rely on ambient light for visibility, do not suffer from luminosity issues, and their optical contrast remains unaffected by pixel size reduction since reflection is governed by polarization of materials at the nanoscale. However, existing reflective display (Electronic Paper (E-paper)) technologies are hampered by significant limitations. Reflective liquid crystal displays, for instance, are constrained by the thickness of the liquid crystal layer, while electrophoretic displays (e.g., Kindle™) are restricted by the size of their capsules (4, 5). To date, no commercially available reflective display technology has achieved high resolutions (>1000 PPI).

Optical metasurfaces have demonstrated the capability to achieve ultrahigh pixel densities > 10,000 PPI (~2.5 μm pixel size), with patterned nanomaterials capable of printing images at resolutions of up to ~100,000 dots per inch, approaching the optical diffraction limit (6, 7, 8). However, most current nanoprinting relies on static materials, such as metals or high refractive index (RI) dielectrics (9, 10, 11). When applied to dynamic display systems, these materials require modulation through micro-light sources, which still suffer from the inherent limitation of electromagnetic reduced intensity as resolution increases (2). In addition, since E-paper does not emit light, interactions between neighboring pixels can alter their optical properties at ultrahigh pixel densities, making it challenging to use conventional RGB subpixel configurations for image display (12, 13, 14).

In recent years, there has been growing interest in integrating dynamic and static materials to explore tunable nanophotonics systems. Particularly in the field of displays, hybrid nanomaterials—combining tunable conjugated polymers or semiconductors as color modulators with metallic nanostructures—have demonstrated the ability to modulate the intensity or the reflected colors of subpixels (15, 16). These technologies significantly enhance the color gamut, reflectivity, and optical contrast of E-paper and enable video display functionality (17, 18, 19, 20). However, due to limitations in structure, materials, and fabrication methods, the pixel sizes of these hybrid nanomaterials remain in the range of tens to hundreds of micrometres, making it challenging to achieve ultrahigh-resolution displays (21,22,23).

Here, we propose a conceptually new E-paper technology, which we name "Retina E-paper", capable of achieving ultrahigh resolutions exceeding 45,000 PPI. The basic principle builds on the hierarchical structuring of building blocks, Mie scattering, and interference between the building blocks. Each pixel has dimensions of ~560 nm, corresponding to a surface area of ~ 1% of the smallest existing LED pixel. This pixel size allows one-to-one correspondence between each subpixel and photoreceptor cells in the human retina, paving the way for the ultimate VR display. The Retina E-paper comprises electrochromic $WO_3$ meta-pixels integrated with a highly reflective substrate (Pt/Al). Its normalized high reflectance (~80%) and optical contrast (~50%) remain unaffected by pixel size reduction, maintaining exceptional visibility even at pixel sizes as small as ~400 nm. To minimize interference between adjacent pixels, we carefully optimized the dimensions and spacing of the primary color meta-pixels, enabling full-color displays by precisely mixing RGB subpixels. In addition, the substrate (Al/Pt) exhibits excellent conductivity. By reducing the lateral distance between the working and counter electrodes to 500 nm and utilizing short-pulse input signals, we achieved >95% optical contrast modulation of the $WO_3$ nano-pixels within 40 ms, supporting video display at 25 Hz. This refresh rate is more than 10 times faster than previously reported fastest $WO_3$-based electrochromic devices (24). It utilizes the small size of the $WO_3$ structure and the fact that the necessary change of the optical density is less than for, e.g., transmission regulation in smart windows to achieve the required optical modulation. Finally, to demonstrate its performance, we used cyan, magenta, and yellow (CMY) color meta-pixels to reproduce the famous painting "The Kiss" by Gustav Klimt and dynamically modulated its colors by electrical control. The surface area of this Retina E-paper is ~1.4 × 1.9 mm², with a resolution of 2100 × 4000 pixels. Despite being ~1/4000 the area of a commercial smartphone display, its resolution is ~2.8 times higher.

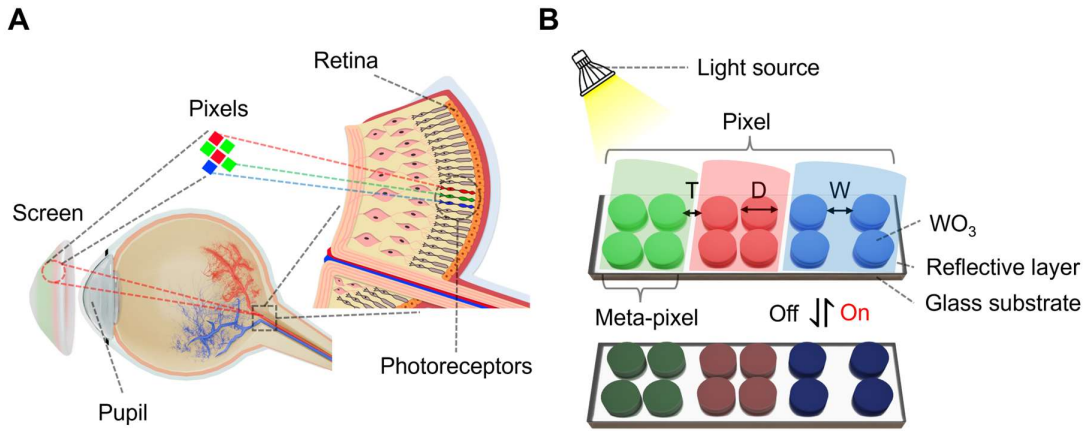

**Fig. 1. Schematic diagram of Retina E-paper.** (**A**) Conceptual illustration of an ultimate virtual reality display: The display dimensions are designed to closely match the human pupil, enabling each pixel to correspond one-to-one with an individual photoreceptor cell in the retina, achieving near-perfect visual fidelity. (**B**) Structural diagram of meta-pixels (sub-pixels): The meta-pixels consist of $WO_3$ nanodiscs and a reflective layer on a glass substrate. By varying the diameter (D) and spacing (W) of the nanodiscs, the meta-pixels can selectively reflect red, green, and blue colors, serving as three subpixels of one pixel for the display. Further tuning of subpixel spacing (T) enables the generation of hybrid colors such as cyan, magenta, and yellow. Since $WO_3$ is electrochromic it can undergo reversible electrochemical reactions, yielding reflectance modulation of the WO3 nanodiscs, allowing for RGB video display.

Fig. 1B illustrates the fundamental structure of the Retina E-paper, composed of electrochromic $WO_3$ metamaterials integrated with a highly reflective (Al/Pt) substrate. $WO_3$ is a semiconductor with a high refractive index RI (~2 to 2.4) in the visible spectrum (25) so that it enables color generation by Mie scattering. By precisely tuning the diameter (D) and spacing

(W) of the nanodiscs, the scattering modes can be adjusted to reflect the primary colors (RGB), forming the subpixels of one pixel of the Retina E-paper. Since the subpixels are made of metamaterial, they are also called meta-pixels. However, optical interactions between these nanodisc-based subpixels can also affect color mixing. To ensure proper additive color blending for display applications, further tuning of the subpixel spacing (T) is required. After patterning the RGB subpixels, the next step is intensity modulation. Since $WO_3$ is an electrochromic material, meta-atom absorption can be dynamically modulated by electrochemical reactions under applied voltage, altering the reflectivity of the subpixels. This capability enables the Retina E-paper to achieve dynamic color modulation for display applications. The nanofabrication process is shown in Fig. S1.

Fig. 2A demonstrates how the reflective colors of $WO_3$ meta-pixels vary at a fixed thickness of 110 nm while varying the nanodisc diameter (D) from 220 nm to 320 nm and the spacing (W) from 100 nm to 200 nm. This range enables the meta-pixels to cover the entire visible spectrum. However, it is essential to note that not all RGB pixels are suitable for subpixels in Retina E-paper. Unlike traditional displays with emissive color light, the reflected color of each subpixel in an ultrahigh resolution E-paper system is influenced not only by its geometry but also by interactions with adjacent pixels (Fig. 2C). Therefore, selecting appropriate RGB pixels and ensuring that their hybrid reflected colors adhere to the principles of additive color mixing is a critical step for achieving full-color displays. On the right side of Fig. 2A, the spectra and corresponding geometries of the selected RGB pixels are presented: R (D = 220 nm, W = 200 nm), G (D = 260 nm, W = 200 nm), and B (D = 260 nm, W = 140 nm). The spectra are normalized to the reflective layer to highlight the $WO_3$ nanodisc color-tuning by structural changes. Unlike emissive displays, where visibility diminishes with pixel size reduction, E-paper technology maintains consistent brightness and reflectivity even at ultrahigh resolutions. As illustrated in Fig. 2B, the red pixel retains its color and reflectivity in both bright-field (BF) and dark-field (DF), even as the size is reduced from 20 μm to 420 nm. To preserve the Mie scattering and grating modes of the nanodiscs, a minimum of four nanodiscs per pixel is required, resulting in minimum pixel sizes of 420 nm for red, 460 nm for green, and 400 nm for blue.

Once the smallest dimensions for the three primary color meta-pixels are determined, the next step is merging them to achieve a full-color display. In Fig. 2C, the grating modes between adjacent subpixels are influenced by the spacing (T) between subpixels, which changes reflected hybrid colors. According to the principles of additive color mixing, the overlap between RGB should produce CMY, respectively. As shown in Fig. 2A, the intermediate region (black dash box) between RGB pixels also contains CMY pixels. As long as carefully designing the spacing between RGB subpixels, the grating modes of adjacent pixels can produce CMY colors, ensuring compliance with the additive color principle. For comparison, Fig. S2 presents several arbitrarily selected combinations of RGB pixels, which fail to reproduce the CMY colors. After carefully selecting RGB pixels and tuning the inter-pixel spacing—300 nm for red-green, 80 nm for blue-red, and 100 nm for green-blue—the desired hybrid colors were successfully generated. The corresponding reflection spectra demonstrate that the reflectance of the hybrid-color pixels matches that of the single-color pixels.

Finally, Fig. 2D presents microscope and SEM images of the merged pixels producing CMY colors. Under high magnification (×100), the arrangement of alternating subpixels along the X-axis to form hybrid colors is clearly visible. It is noteworthy that the Mie scattering mode of individual nanodiscs is primarily determined by their size, while the grating mode of the subpixel arrays in the X-direction governs the generation of reflective mixed colors.

As an electrochromic material, $WO_3$ exhibits an electrically tunable refractive index (n) and extinction coefficient (k) across the visible spectrum (400–700 nm). In the oxidized (color) state, the refractive index varies from ~2.38 to 2.14, with k values <0.01. In the reduced (black) state, n decreases from ~2.25 to 1.95, while k increases significantly, k > 0.4 (Fig. S3). Fig. 3A presents the experimental setup to electrochemically control the color states of $WO_3$ meta-pixels. The electrolyte consists of 1 M LiClO₄ in acetonitrile, and the RGB pixel size is 350 μm. Metallic electrodes (Pt/Al) were employed to minimize potential drops. Notably, we designed a lateral electrode configuration with a narrow 500 nm gap between the working and counter electrodes, enhancing the local electric field and significantly improving the switching speed (26).

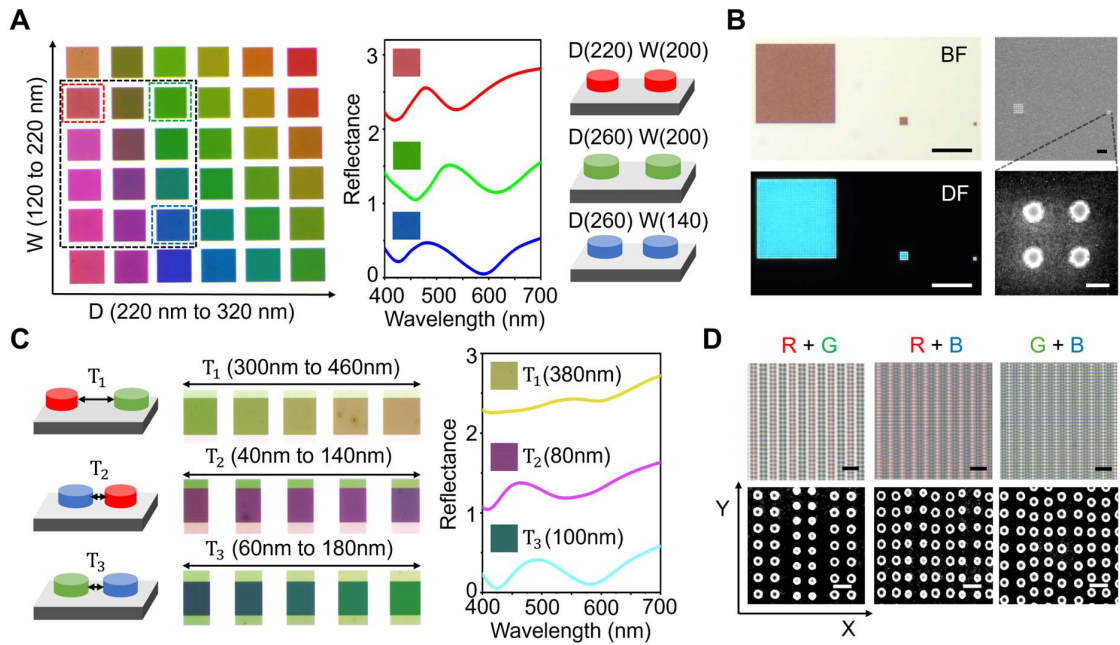

**Fig. 2. Design and characterization of $WO_3$ meta-pixels.** (**A**) Meta-pixel design: Left: Tuning the diameter (D) and spacing (W) of $WO_3$ nanodiscs achieves a diverse color palette. The dashed box highlights selected red, green, and blue (RGB) pixels along with their intermediate regions, which contain cyan, magenta, and yellow (CMY) pixels. Middle: Reflectance spectra of the selected RGB pixels. Right: Corresponding D and W values for the chosen RGB pixels. (**B**) Microscopic and structural characterization: Left: Bright-field (BF) and dark-field (DF) microscope images of a red pixel with feature sizes of 20 μm, 2 μm, and 420 nm, captured under ×100 high magnification. Scale bar: 10 μm. Right: Scanning electron microscope (SEM) images of red pixels with 2 μm and 420 nm. Scale bars: 2 μm and 200 nm, respectively. (**C**) Color mixing by subpixel arrangement: Left and Middle: Reflective color varies as a function of the spacing (T) between adjacent RGB subpixels. Right: Reflectance spectra of hybrid yellow, magenta, and cyan pixels, corresponding to optimized subpixel spacing. (**D**) High-resolution color imaging: Top: Bright-field microscope (×100) images of hybrid yellow, magenta, and cyan pixels. Bottom: SEM images of the corresponding hybrid pixels. Scale bars: 1 μm (microscope) and 500 nm (SEM).

Fig. 3B illustrates the measured normalized reflectance modulation of RGB meta-pixels in the ON/OFF states. Since both Mie scattering and grating modes are influenced by variations in the refractive index of the surrounding environment, the optimized nanodisc dimensions for the RGB meta-pixels are: R (D = 300 nm, W = 120 nm), G (D = 280 nm, W = 80 nm), and B (D = 260 nm, W = 40 nm). A clear distinction is observed when comparing the meta-pixel reflectance in air versus in electrolytes. Specifically, the reflectance in electrolytes is notably higher, while scattering in the red-light region is significantly suppressed. This effect arises because Mie scattering depends on the relative refractive index $m = \frac{n_{particle}}{n_{env}}$, where $n_{particle}$ and $n_{env}$ are the refractive indices of the WO₃ nanodiscs and the surrounding environment, respectively. When $n_{particle}$ is close to $n_{env}$, the scattering is drastically reduced, making the material appear more transparent and desaturated in color. Consequently, the color contrast of WO₃ meta-pixels in the electrolytes is less saturated than in the air. However, despite the ultra-thin ~100 nm thickness of the WO₃ nanodiscs, the optical contrast remains ~50%, significantly outperforming most of the same thick planar WO₃ electrochromic devices (27). This enhancement is due to the high refractive index of WO₃, concentrating the electric field of incident light within the nanodiscs. After switching to the dark state, most of the incident light is absorbed, as confirmed by numerical simulations of red subpixel on/off states (Fig. 3B right).

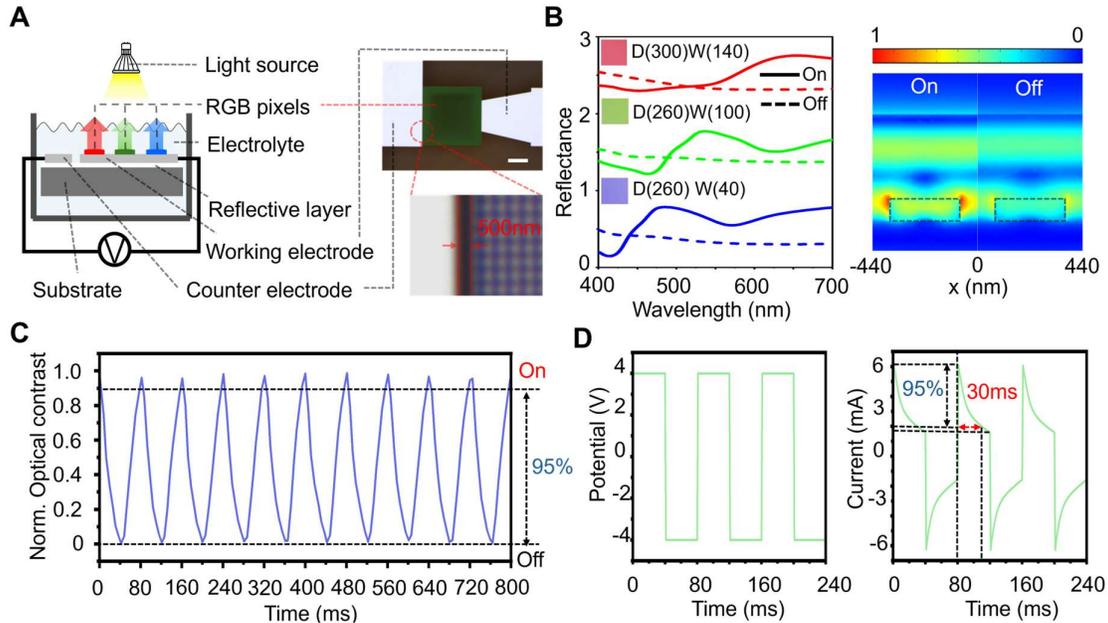

**Fig. 3. Electrochemical modulation of WO₃ meta-pixels.** (**A**) Experimental setup for electrochemical characterization: Left: Cross-sectional schematic diagram illustrating the setup used to characterize the optical properties of WO₃ meta-pixels under electrochemical control. Right: Microscope images of the characterized sample show a 500 nm gap between the working and counter electrodes, which enhances the local electric field for improved switching performance. Scale bar: 100 μm. (**B**) Optical response of electrochemically-controlled RGB meta-pixels: Left: Reflectance spectra of the selected red, green, and blue meta-pixels in their electrochemically modulated on/off states, demonstrating dynamic tunability. Right: Numerical simulation of the reflectance modulation of red meta-pixels at a 650 nm incident wavelength, validating that most of the electric field is concentrated within the WO₃ nanodiscs. (**C**) Switching speed characterization: By applying a short pulse voltage signal, 95% normalized optical contrast is achieved within 40 ms, demonstrating video-rate display applications. (**D**) Electrical response corresponding to optical modulation: Left: Input voltage signal of ±4 V with a 40 ms pulse duration. Right: The corresponding current response, where 95% of the current variation is achieved within 30 ms, confirms the fast ion transport dynamics in the WO₃ nanodisc structure.

Benefiting from the strong local electric field between the closely spaced working and counter electrodes, as well as the ultra-thin amorphous WO₃ nanodiscs, yielding fast ion insertion (28), the electrochemically tunable meta-pixels achieve an exceptionally fast switching time of only 40 ms in order to reach >95% of the total optical contrast modulation. Movie S1 demonstrates the rapid ON/OFF switching of the blue meta-pixel, confirming they support video display. Fig. 3C shows 10 cycles of reflectance variation in ±4V pulse input signals. The normalized optical contrast is calculated by $\frac{R-R_{min}}{R_{max}-R_{min}}$, where $R$, $R_{min}$ and $R_{max}$ are the real-time, minimum and maximum reflectance of the WO₃ meta-pixels, respectively, to clearly illustrate the optical contrast change. The measured optical contrast is shown in Fig. S7. Notably, the operating voltage is comparable to the solid-state two-electrode WO₃ electrochromic systems (23). However, due to the significantly enhanced switching speed (>×65 faster), precise pulsed voltage control—rather than a constant bias—can be employed to minimize energy consumption and mitigate side effects. Furthermore, in practical display applications, since not all pixels undergo 100% intensity changes in a typical video frame (<10% variation per frame), the effective response time will be significantly shorter than 40 ms (19).

To further demonstrate the display capabilities of the Retina E-paper, we reproduced the renowned painting "The Kiss" by Gustav Klimt by fabricating meta-pixels. We chose this painting for its complex patterns and diverse colors. Since our substrate is a highly reflective material analogous to a white canvas, we employed the CMY sub-pixels and used subtractive color mixing to render the image. It is important to note that the patterned image only demonstrates the display capability of WO₃ meta-pixels. For the display application, a TFT array should be employed to independently control the reflectance of each pixel, while the background should be set to black. The image rendering should follow the additive color principle using RGB subpixels, as illustrated in Fig. 2.

Fig. 4A (left panel) illustrates the nanodisc diameters and periodicities for the CMY meta-pixels alongside their corresponding reflection spectra, which show similar reflectance as RGB pixels. The right panel presents the merged hybrid color meta-pixels, where the spacing between subpixels are B ($T_1$ = 100 nm), R ($T_2$ = 60 nm), and G ($T_3$ = 60 nm). Notably, except for similar reflectance to RGB meta-pixels, the CMY system also ensures that the intermediate regions contain RGB pixels (Fig. S4). The optimized nanodisc dimensions for CMY pixels are C (D = 260 nm, W = 160 nm), M (D = 240 nm, W = 100 nm), and Y (D = 180 nm, W = 180 nm).

Fig. 4B directly compares the Retina E-paper and a commercial mobile display (iPhone 15) in terms of both physical dimensions and display resolution. While the iPhone screen measures 147.6 mm × 71.6 mm, the Retina E-paper is only 1.9 mm × 1.4 mm, amounting to merely

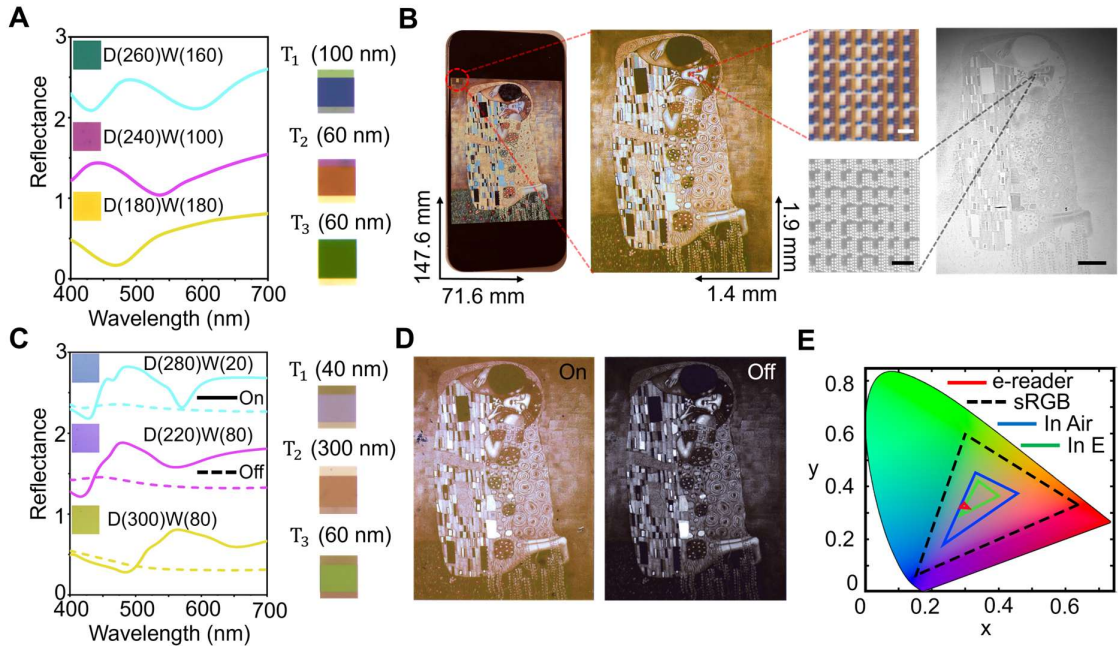

**Fig. 4. Characterization of Retina E-Paper display performance.** (**A**) Optical properties of CMY meta-pixels: Left: The reflectance spectra of the selected cyan, magenta, and yellow (CMY) meta-pixels demonstrate the spectral response. Right: Photographs of the red, green, and blue (RGB) pixels with optimized adjacent subpixel spacing for improved hybrid color and display fidelity. (**B**) High-resolution display of "The Kiss" on Retina E-Paper vs. iPhone 15: Photographs comparing the display of "The Kiss" on an iPhone 15 and Retina E-paper. The surface area of the Retina E-paper is ~ 1/4000 times smaller than the iPhone 15. SEM and microscope images confirm that the displayed colors are generated by precisely arranged CMY subpixels. Scale bars: 2 µm, 2 µm, 200 µm. (**C**) Electrochemical modulation of CMY meta-pixels: Left: The Reflectance spectra of CMY meta-pixels in the electrochemically modulated on/off states demonstrate dynamic tunability. Right: Photographs of RGB hybrid colors in an electrolyte environment with optimized adjacent CMY subpixel spacing. (**D**) Electrochemical display of "The Kiss" by Retina E-Paper: Photographs showing the display of "The Kiss" on Retina E-paper in the on/off states, demonstrating reversible color modulation under electrochemical adjusting. (**E**) Comparison of color gamut coverage: CIE chromaticity diagram comparing the color coverage of an emissive display (sRGB), a commercial electrophoretic display (e-reader), and the Retina E-paper in both air and electrolyte environments. The results highlight the color performance of the Retina E-paper relative to conventional display technologies.

~1/4000 the area of the smartphone display. Despite this minuscule size, the Retina E-paper 1achieves 4000 × 2100 resolution, approximately 2.8 times higher than the 2556 × 1179 resolution of the smartphone display. Since the Retina E-paper does not connect to thin-film transistor (TFT) arrays to individually adjust each subpixel, it achieves color rendering solely by the precise geometric design of CMY meta-pixels, as detailed in Fig. S5. Due to the inherent challenges of accurately controlling the reflectance of subpixels and the narrower color gamut compared to emissive displays, the perceived color saturation of the Retina E-paper is lower than that of the iPhone 15. However, this is the first demonstration of full-color imaging achieved by three primary color meta-pixels at such an ultrahigh resolution. With an average pixel size of only ~560 nm, the display reaches an unprecedented >45,000 PPI, surpassing the resolution requirements for ultimate VR displays. High-magnification microscope (×100) and SEM images (Fig. 4B right) further confirm the well-ordered CMY meta-pixel arrangement and the vibrant color rendering, validating the ultrahigh resolution of the Retina E-paper.

To evaluate its electrically tunable color performance, we again reproduced "The Kiss" using CMY meta-pixels in an electrolyte environment. Fig. 4C presents the corresponding reflection spectra of CMY pixels in their color and dark states. To maintain the presence of RGB pixels in the intermediate regions (Fig. S6), the dimensions of the CMY meta-pixels were further optimized: C (D = 280 nm, W = 20 nm), M (D = 220 nm, W = 80 nm), Y (D = 220 nm, W = 80 nm). The merged RGB subpixel spacing was adjusted accordingly: B ($T_1$ = 40 nm), R ($T_2$ = 300 nm), G ($T_3$ = 60 nm). Fig. 4D showcases the photos of "The Kiss" under color and dark states in the electrolyte. Due to the weaker Mie scattering of $WO_3$ nanodiscs in electrolyte compared to air, the displayed colors appear less saturated, with a noticeable reduction in extinction within the red region, resulting in an overall red-shifted color. However, the system exhibits a distinct reflectance modulation between ON and OFF states, highlighting its potential for dynamic display applications.

Finally, Fig. 4E compares the color gamut coverage of commercial emissive displays, the Retina E-paper in both air and electrolyte and the commercial color electrophoretic display. Although the color gamut of Retina E-paper remains narrower than emissive displays, its performance in both the air and electrolyte significantly surpasses commercial color e-paper (29).

Nowadays, more than 80% of the information people perceive is by visual signals (30). With the development of Internet of Things (IoT) technology and increasing information transfer speeds, the demand for next-generation visual display technologies keeps growing. Retina E-paper not only reaches the theoretical resolution limit of human vision but also offers exceptional visibility. It enables full-color video display while maintaining high reflectivity and optical contrast, which is promising for realizing ultimate VR displays. Despite these advantages, the Retina E-paper still requires further optimization of color gamut coverage, refresh rate, and lifetime. Moreover, its ultra-high resolution necessitates the development of ultra-high resolution TFT arrays for independently controlling each pixel. In the future, we anticipate significant advancements in this field and firmly believe that the evolution of the Retina E-paper will ultimately influence everyone.